\documentclass[preprint,aps]{revtex4}
\usepackage{graphicx}
\usepackage{amssymb}
\newcommand{\be}{\begin{eqnarray}}
\newcommand{\ee}{\end{eqnarray}}
\usepackage{hyperref}
\usepackage{subfigure}
\usepackage{cancel} 
\newcommand{\nn}{\nonumber}

\usepackage{tikz}

\begin{document}

\title{Generalized Parton Distributions of the Photon with Helicity Flip}

\author{\bf Asmita Mukherjee , Sreeraj Nair and Vikash Kumar Ojha}

\affiliation{ Department of Physics,
Indian Institute of Technology Bombay,\\ Powai, Mumbai 400076,
India.}
\date{\today}

\begin{abstract}
We present a calculation of the generalized parton distributions (GPDs) of the
photon when the helicity of the initial photon is different from the final
photon. We calculate the GPDs using overlaps of photon light-front wave
functions (LFWFs) at leading order in electromagnetic  
coupling $\alpha$ and zeroth order in the strong coupling $\alpha_s$, when
the momentum transfer is purely in the transverse direction. These involve
a contribution of orbital angular momentum of two units in the LFWFs.  
We express these GPDs in the impact parameter space. 

\end{abstract}
\maketitle

{ \bf{Introduction}}

\vspace{0.2in}
Generalized parton distributions (GPDs) of the nucleon are unified objects giving a
wide range of information on nuclear structure and spin \cite{rev}. 
These are
non-perturbative objects appearing in the factorized amplitude of exclusive
processes like deeply virtual Compton scattering (DVCS) and meson production;
 and can be expressed as an off-forward matrix element of light-cone bilocal operators. 
In \cite{pire} the amplitude of the DVCS process on a photon target $ \gamma^*(Q) \gamma \to \gamma
\gamma$ at high $Q^2$ is written in terms of photon GPDs. These photon GPDs
were calculated at leading order in electromagnetic coupling  $\alpha$ and zeroth                           
order in the strong coupling $\alpha_s$  and upto leading logs; in the
kinematical limit that there is no momentum transfer in the transverse
direction. In fact the parton content of the photon is known to play an
important role in high energy scattering processes. The parton distributions
of the photon are now well understood both theoretically and experimentally
\cite{photon}.
On the other hand, the GPDs and generalized distribution amplitudes (GDAs)
of the photon \cite{gda} are much less
investigated objects. In a  couple of recent works \cite{ours1,ours2}, we 
extended the calculation of photon GPDs in the more general kinematics when the momentum
transfer has both transverse and longitudinal components. We have
developed an overlap representation using the light-front wave function of
the photon. We also showed that the impact parameter space interpretation of
the photon GPDs give a 3D position space description of them. In another
recent work \cite{ter}, GPDs of the photon have been used to investigate analyticity
properties of DVCS amplitudes and related sum rules for the GPDs. 

As we know, in the DVCS process $eP \to e \gamma P$, the helicity of the
proton may or may not flip due to the scattering. When the proton helicity
is flipped, the DVCS amplitude is parametrized in terms of the GPD $E$
\cite{rev}.
This flip requires non-zero orbital angular momentum in the overlapping
light-front wave functions (LFWFs) and is not possible unless there is
non-zero momentum transfer in the transverse direction. For a transversely
polarized nucleon, this gives a distortion of the parton distributions in
the transverse position or impact parameter space \cite{burkardt}.  In two 
previous
articles, we calculated  the impact parameter space representations of the photon
GPDs when the helicity of the photon is not flipped. In this work, we
calculate the GPDs that involve helicity flip of the photon and represent
them in impact parameter space. Like the proton, these involve overlaps of
LFWFs of the photon, with non-zero orbital angular momentum (OAM). The
corresponding parton distributions in the impact parameter space show 
distortions related to the orbital angular momentum of the LFWFs.       
 
\vspace{0.2in}   
{\bf{GPDs of the photon with helicity flip}}
\vspace{0.2in}

The GPDs of the photon can be expressed as the following off-forward matrix
elements \cite{ours1,ours2}\\
\be
F^q=\int {dy^-\over 8 \pi} e^{-i P^+ y^-\over 2} \langle \gamma(P'),\lambda' \mid
{\bar{\psi}} (0) \gamma^+ \psi(y^-) \mid \gamma (P),\lambda \rangle
\label{def}
\ee
\be
\tilde{F^q}=\int {dy^-\over 8 \pi} e^{-i P^+ y^-\over 2} \langle \gamma(P'),\lambda' \mid
{\bar{\psi}} (0) \gamma^+  \gamma_5 \psi(y^-) \mid \gamma (P),\lambda \rangle
\label{def2}
\ee \\
here $\mid \gamma(P), \lambda \rangle$ is the (real) photon target state of
momentum $P$ and helicity $\lambda$. We work in the light-front gauge
$A^+=0$. We use the standard LF coordinates $P^\pm = P^0 \pm P^3,~ y^\pm = 
y^0 \pm y^3$.
Since the target photon  is on-shell, $P^+ P^- -{P^\perp}^2 = 0$, 
the momenta of the initial and final photon in the most general case of
momentum transfer are given by:
\begin{eqnarray}
P&=&
\left(\ P^+\ ,\ {0^\perp}\ ,\ 0\ \right)\ ,
\label{a1}\\
P'&=&
\left( (1-\zeta)P^+\ ,\ -{\Delta^\perp}\ ,\ 
{{\Delta}^{\perp 2} \over (1-\zeta)P^+}\right)\ ,
\end{eqnarray}
The four-momentum transfer from the target is
\begin{eqnarray}
\label{delta}   
\Delta&=&P-P'\ =\
\left( \zeta P^+\ ,\ {\Delta^\perp}\ ,\
{t+{\Delta^\perp}^2 \over \zeta P^+}\right)\ ,
\end{eqnarray}
where $t = \Delta^2$ and $\zeta$ is called the skewness variable. In addition, overall energy-momentum
conservation requires $\Delta^- = P^- - P'^-$, which connects 
${\Delta^\perp}^2$, $\zeta$, and $t$ according to
\begin{equation}
 (1-\zeta) t  = -{\Delta^\perp}^2 .
 \label{tzeta}
\end{equation}

In order to calculate the above matrix element, we use  the Fock space expansion of
the photon state, which can be written as \cite{ours1} 
\be
\mid \gamma(P),\lambda \rangle &=& \sqrt{N} \Big [ a^\dagger(P, \lambda) \mid 0
\rangle + \sum_{\sigma_1, \sigma_2} \int \{dk_1\} \int \{ dk_2\} \sqrt{2{(2
\pi)}^3 P^+} \delta^3 (P-k_1-k_2)\nonumber\\&&~~~~ \phi_2(k_1,k_2,\sigma_1,
\sigma_2)
b^\dagger(k_1, \sigma_1) d^\dagger(k_2, \sigma_2) \mid 0 \rangle \Big ]
\ee 
where $\sqrt{N}$ is the normalization of the state; which in our
calculation we can take as unity as any correction to it contributes at 
higher order in $\alpha$.  $\{ dk\}= \int {dk^+ d^2 k^\perp\over \sqrt{2 {(2
\pi)}^3 k^+}}$, $\phi_2$ is the
two-particle ($q \bar{q}$) light-front wave function (LFWF) and $\sigma_1$
and
$\sigma_2$ are the helicities of the quark and antiquark. The wave function   
can be expressed in terms of Jacobi momenta $x_i={k_i^+\over P^+}$ and
$q_i^\perp=k_i^\perp-x_i P^\perp$. These obey the relations $\sum_i x_i=1,
\sum_i q_i^\perp=0$. The Lorentz boost invariant two-particle LFWFs are given by     
${\psi_2(x_i,q_i^\perp)=\phi_2 \sqrt{P^+}}$. $\psi_2(x_i, q_i^\perp)$
can be calculated order by order in perturbation theory.
The two-particle LFWFs for the photon are given by
\be
\psi_{2 s_1, s_2}^\lambda(x,q^\perp) &=& {1\over m^2-{m^2+{(q^\perp)}^2
\over x (1-x)}} {e e_q\over \sqrt{ 2 {(2 \pi)}^3}} 
\chi^\dagger_{s_1} \Big [
{(\sigma^\perp \cdot q^\perp)\over x} \sigma^\perp \nonumber\\&&- \sigma^\perp
{(\sigma^\perp \cdot q^\perp)\over 1-x} -i {m \over x (1-x)} \sigma^\perp
\Big ] \chi_{-s_2} \epsilon^{\perp *}_{\lambda} 
\ee
where $m$ is the mass of $q(\bar{q})$. $\lambda$ is  the 
helicity of the photon and $s_1,s_2$ are the helicities of the $q$ and ${\bar q}$ respectively.
We have used the two-component form of light-cone field theory \cite{two}, namely
the component $A^-$ of the photon field is constrained in the gauge $A^+=0$
and can be eliminated from the theory. So one has only the transverse
components of the photon field $A^\perp$. Likewise, the 'bad' component of
the fermion field $\psi^{(-)}$ is eliminated using constraint equation and $
\psi^{(+)}$ is written in terms of two-component spinors, $\chi_{s}$
\cite{two}.

The GPDs can be written in terms of the overlaps of the LFWFs as follows :
\be 
F^q &=& \int d^2 q^\perp dx_1 \delta(x-x_1) \psi_2^{*\lambda'}(
x_1,q^\perp-(1-x_1) 
\Delta^\perp
)\psi_2^\lambda (x_1, q^\perp)\nonumber\\&&~~-\int d^2 q^\perp dx_1 
\delta(1+x-x_1) \psi_2^{*\lambda'}(x_1,q^\perp+x_1 \Delta^\perp
)\psi_2^\lambda (x_1, q^\perp) 
\label{lfwf}
\ee

We calculate the photon GPDs using overlaps of light-front wave
functions. We take the momentum transfer to be purely in the transverse
direction, unlike \cite{pire}, where the momentum transfer was taken purely
in the light-cone (plus) direction. GPDs in this case can be expressed in
terms of diagonal 
(particle number conserving) overlaps of LFWFs. When there is non-zero momentum  
transfer in the longitudinal direction, there are off-diagonal particle
number changing overlaps as well, similar to the proton GPDs \cite{overlap}.

The transverse polarization vector of the photon can be written as :
\be 
\epsilon^\perp_\pm = \frac{1}{\sqrt{2}}(\mp1,-i)
\ee

We extract the GPD that involves a helicity
flip of the target photon from the non-vanishing coefficient of the
combination $(\epsilon^1_{+1} \epsilon^{1*}_{-1}+\epsilon^2_{+1} 
\epsilon^{2*}_{-1})$. The corresponding
GPD without a helicity flip of the photon contains a leading logarithmic term at
leading order in $\alpha$ and zeroth order in strong coupling constant and
has been discussed in two previous articles \cite{ours1,ours2}.
 The GPD with helicity flip is given by :     	
\be
E_1 = \frac{\alpha e_q^2}{2 \pi^2}x (1-x) \Big[ I_1 - (1-x)I_2 \Big].
\ee
The integrals $I_1$ and $I_2$ are given by :
\be
I_1 = \int d^2q^\perp \frac{((q^1)^2 -(q^2)^2)} {D_1 D_2}
\hspace{1cm}I_2 = \int d^2q^\perp \frac{(q^1\Delta^1 - q^2\Delta^2)} {D_1
D_2};
\nn 
\ee
where $q^1$ and $q^2$ are the $x$ and $y$ components of $q^\perp$ and
$\Delta^1$ and $\Delta^2$ are the $x$ and $y$ components of $\Delta^\perp$
respectively. The denominators are given by :

\be 
D_1 = {(q^\perp)}^2 -m^2 x (1-x) +m^2 \nonumber\\
D_2 = 
{(q^\perp)}^2 +(1-x)^2 
{(\Delta^\perp)}^2 -2 q^\perp \cdot \Delta^\perp (1-x)- m^2 x (1-x) +m^2. 
\ee

In order to simplify the above expression we use the formula \cite{rad}
\be
\frac{1}{A^k} = \frac{1}{\Gamma(k)} \int_0^\infty \alpha^{k-1} 
e^{-\alpha A} d\alpha .
\ee

The integrals can be written in the form :
\be
I_1 =  ({(\Delta^1)}^2-{(\Delta^2)}^2)\pi (1-x)^{2} \int_0^{1} dq\hspace{0.2cm}
\frac{(1-q)^2}{B(q)};~~~
I_2= ({(\Delta^1)}^2-{(\Delta^2)}^2) \pi (1-x) \int_0^{1} dq\hspace{0.2cm} 
\frac{(1-q)}{B(q)}; 
\nn \ee
where
\be
B(q) = m^2\Big(1-x (1-x)\Big) +q (1-q) (1-x)^2 
{(\Delta^\perp)}^2.
\ee
\begin{figure}
\begin{minipage}[c]{0.99\textwidth}
\tiny{(a)}\includegraphics[width=7cm,height=6cm,clip]{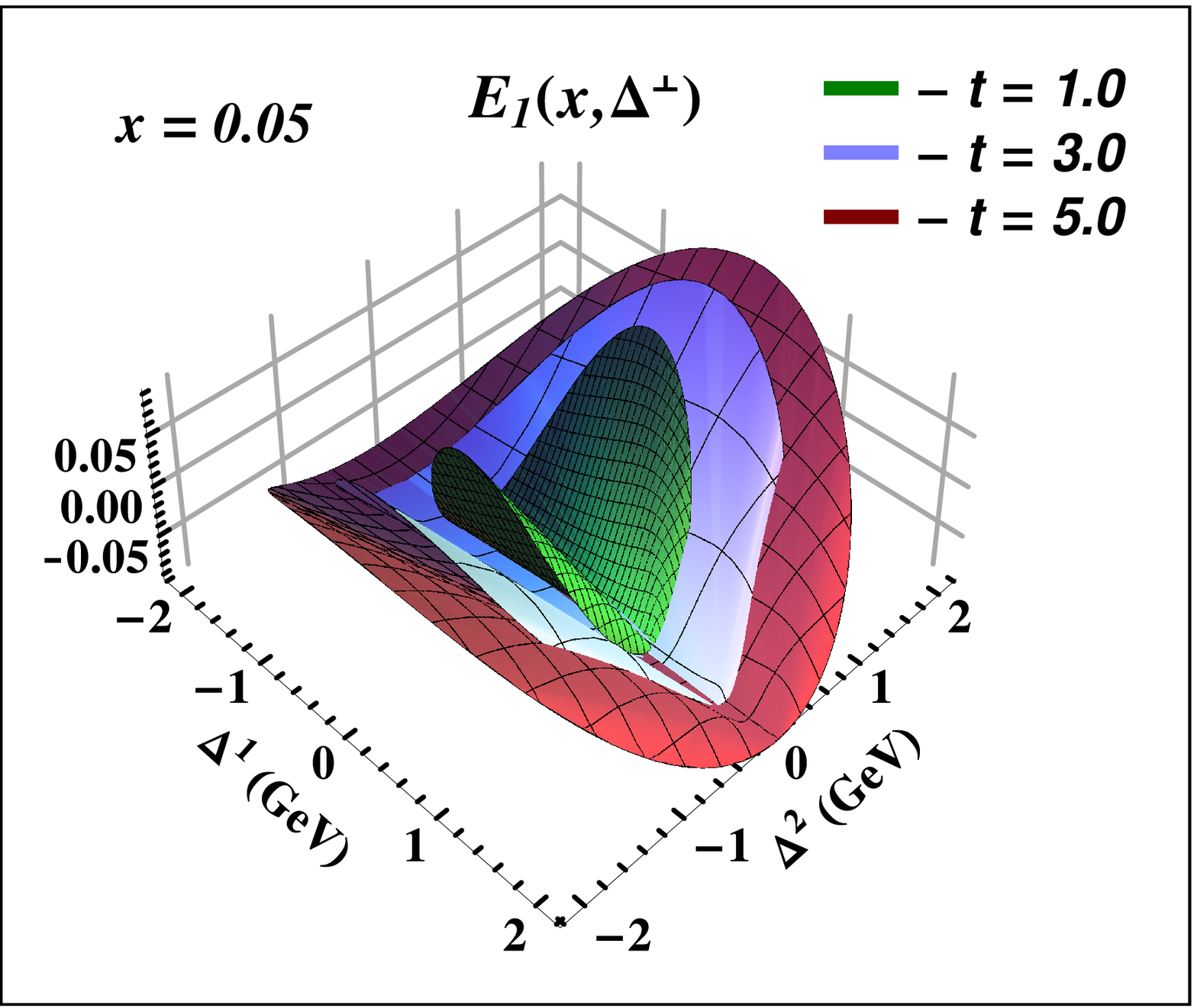}
\hspace{0.1cm}
\tiny{(b)}\includegraphics[width=7cm,height=6cm,clip]{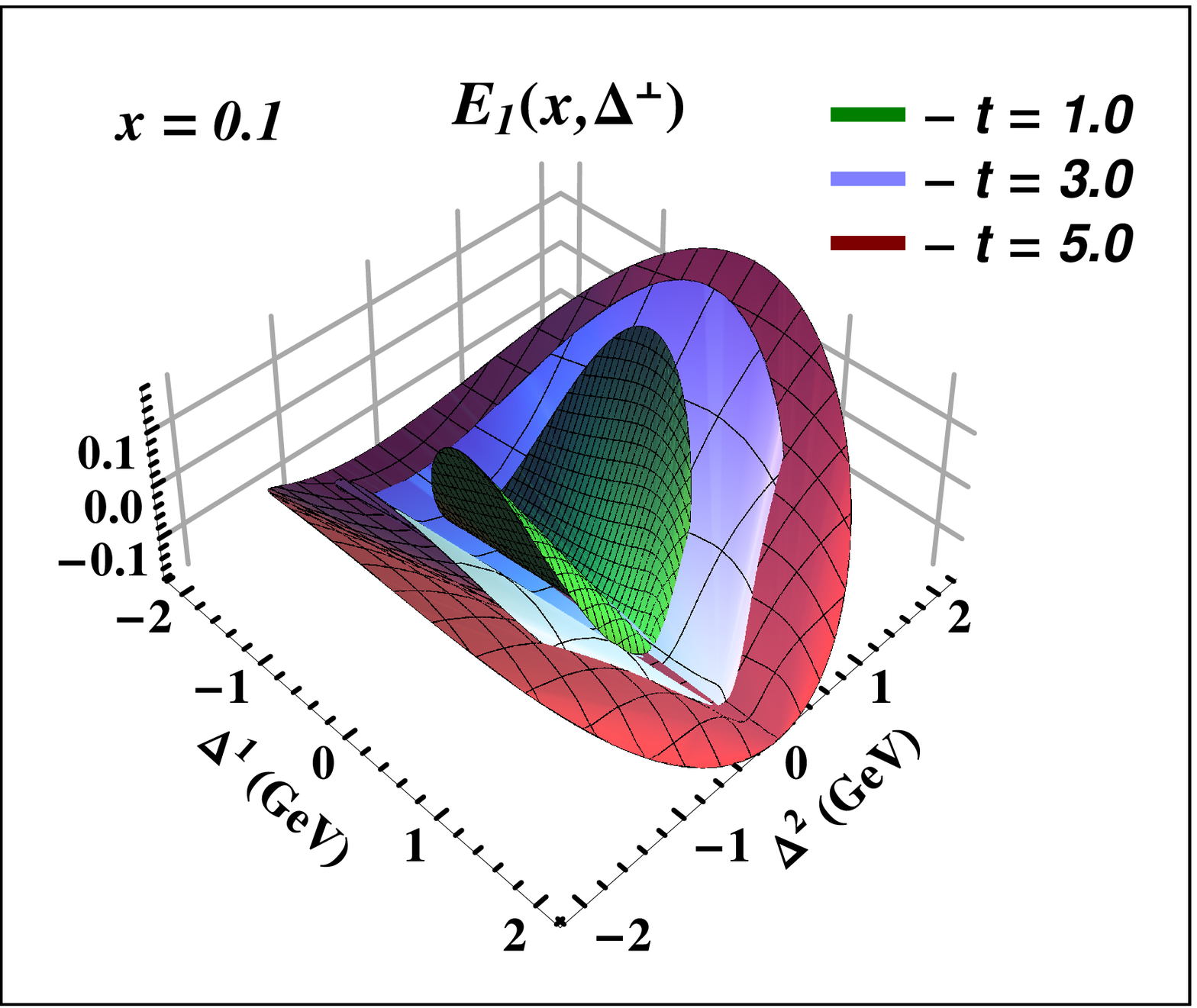}
\end{minipage}
\caption{\label{fig1}(Color online) Plots of  $E_1(x,\Delta^\perp)$ vs 
$\Delta^1,\Delta^2$ for fixed values of $x$ and different $t$. $t$ is in
${\mathrm{GeV}}^2$. The innermost surface is for the smallest value of $-t$.} 
\end{figure}
So we have; 
\be
E_1 = \frac{\alpha e_{q}^2}{2 \pi}x (1-x)^{3} ({(\Delta^1)}^2-
{(\Delta^2)}^2)
\Big[ \int_0^{1} \frac{dq}{B(q)}\hspace{0.2cm}
((1-q)^2-(1-q)) \Big]. 
\ee 

The above has the expected quadrupole structure coming from $
({(\Delta^1)}^2-{(\Delta^2)}^2)$. As the photon is a spin one particle, in
order to flip its helicity, the overlapping light-front wave functions
should have a difference of orbital angular momentum of two units, which
manifests itself in the quadrupole structure. This is in accordance with a
similar observation for the helicity-flip GPD $E$ for the proton, which
needs overlapping LFWFs of orbital angular momentum $\pm 1$ unit
\cite{overlap,spin}. 

\begin{figure}
\begin{minipage}[c]{0.99\textwidth}
\tiny{(a)}\includegraphics[width=7cm,height=6cm,clip]{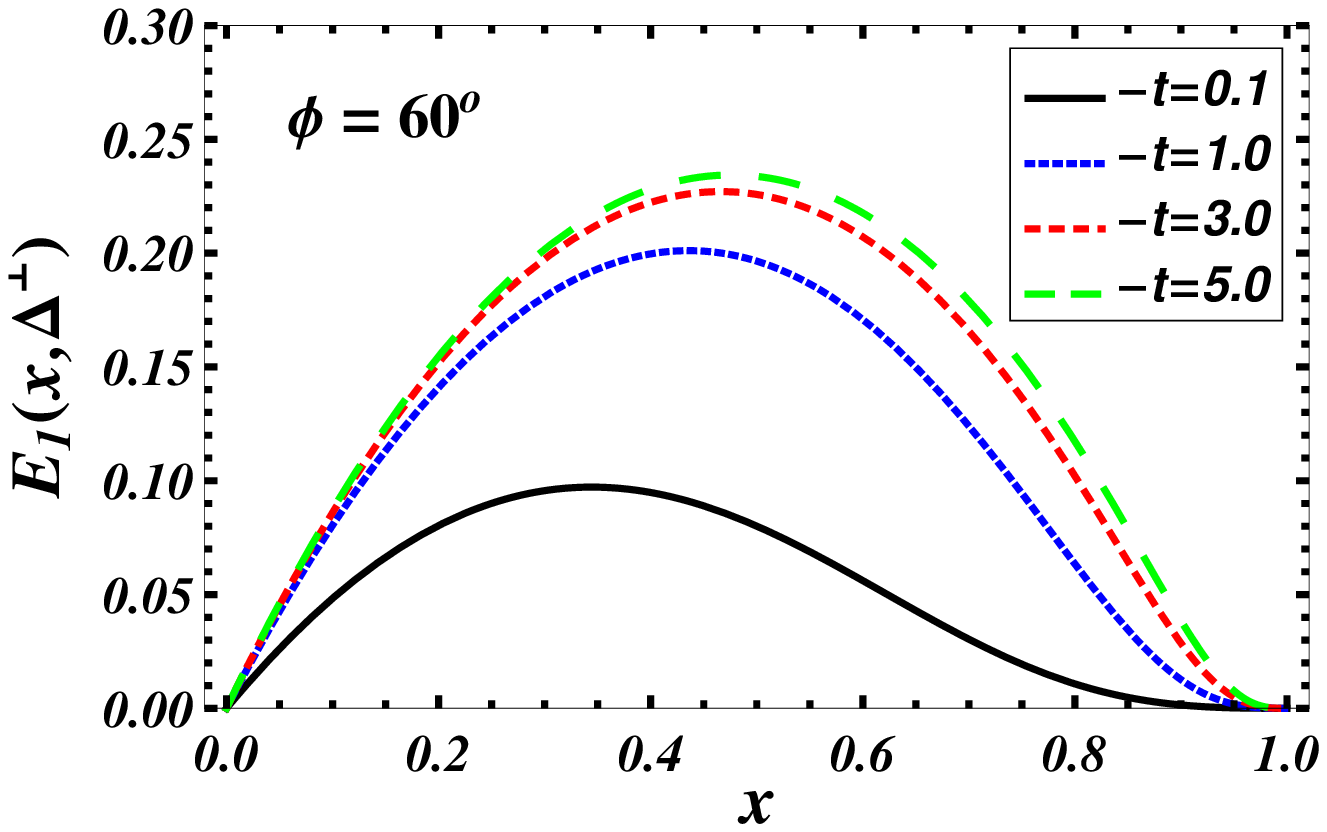}
\hspace{0.1cm}
\tiny{(b)}\includegraphics[width=7cm,height=6cm,clip]{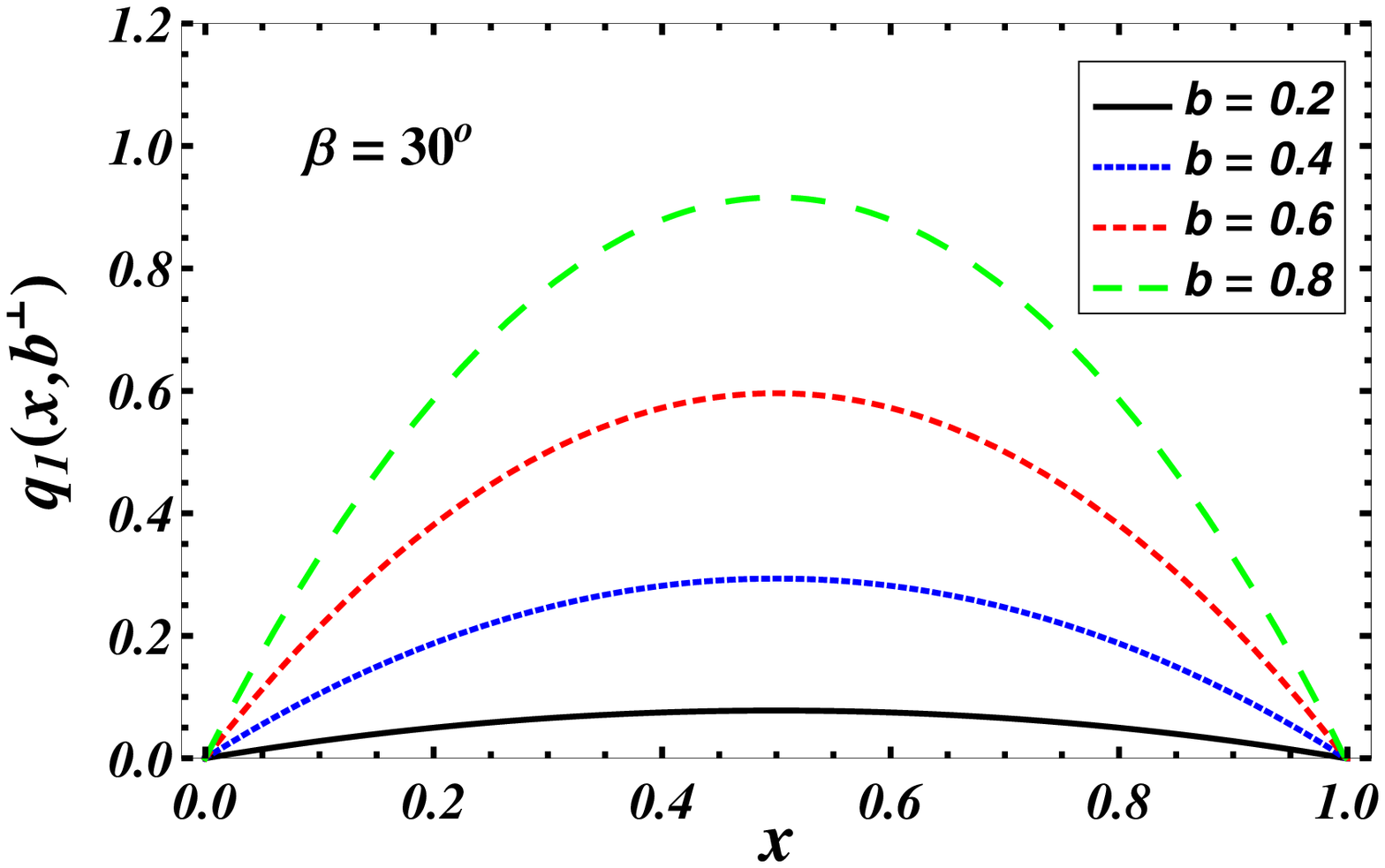}
\end{minipage}
\caption{\label{fig2}(Color online) Plots of  (a) $E_1 (x,\Delta^\perp)$ vs 
$x$ for different values of $t$ in $\mathrm{GeV}^2$ and (b) $q_1(x,b^\perp)$ vs.
$x$ for different values of $b$ in $\mathrm{GeV}^{-1}$.
$\phi={\mathrm{tan}}^{-1}{\Delta^2\over \Delta^1}$ and $\beta=
{\mathrm{tan}}^{-1}{b^2\over b^1}$.}
\end{figure}

From the off-forward matrix element $\tilde F_q$ we extract the GPDs 
that flip the helicity of the photon by calculating the coefficient of the 
combination $(\epsilon_{+1}^{1} \epsilon_{-1}^{2*}  
+ \epsilon_{+1}^{2} \epsilon_{-1}^{1*}) $
which gives:\\
\be
\tilde{E_1}  = \frac{\alpha e_{q}^2}{4 \pi^2}(x-x') \Big[
\tilde  I_1 - (1-x') \tilde I_2 \Big]
\nn  \ee
   
\be
\tilde  I_1 =  \pi  \int_0^{1} dq\hspace{0.2cm}
\Big((\Delta_{1}^{2}-\Delta_{2}^{2})\frac{(1-q)^2}{B(q)}(1-x')^{2}\Big)
\nn \ee
\be
\tilde  I_2= \pi \int_0^{1} dq\hspace{0.2cm} \hspace{0.2cm}
(\Delta_{1}^{2}-\Delta_{2}^{2})\frac{(1-q)}{B(q)} (1-x')
\nn \ee
   
\be
\tilde{E_1} = \frac{\alpha e_{q}^2}{2 \pi}(x-x') (1-x')^{2}
(\Delta_{1}^{2}-\Delta_{2}^{2})
\Big[ \int_0^{1} \frac{dq}{B(q)}\hspace{0.2cm}
 (-q)(1-q) \Big]   
 \label{e3t} \ee \\

Here $x'$ is the longitudinal momentum fracion of the quark in the final
photon LFWF. As we have taken the momentum transfer to be purely in the
transverse direction, $x'=x$ and $\tilde E_1=0$. $E_1$ and $\tilde E_1$ are
only two independent structures that cause helicity flip of the photon.
Other helicity-flip GPDs that can be constructed from other combinations of
the polarization vectors can be  related to these by phase change in the
$\Delta^\perp$ plane. However a proper counting of the photon GPD (both
helicty non-flip and helicity flip) can only be done in a formal parametrization
of Eqs.  (\ref{def}) and (\ref{def2}).

Like the GPD $E$ of a spin $1/2$ particle for example  a dressed electron/quark
\cite{quark}, the helicity flip photon GPD has no
logarithmic term depending on the hard scale of the process $Q^2$, which is
the virtuality of the probing photon.  
Starting from the expressions of photon GPDs, we define the 
parton distributions \cite{burkardt} with  the helicity flip of the photon  in transverse impact 
parameter space as:

\begin{figure}
\begin{minipage}[c]{0.99\textwidth}
\tiny{(a)}\includegraphics[width=7cm,height=6cm,clip]{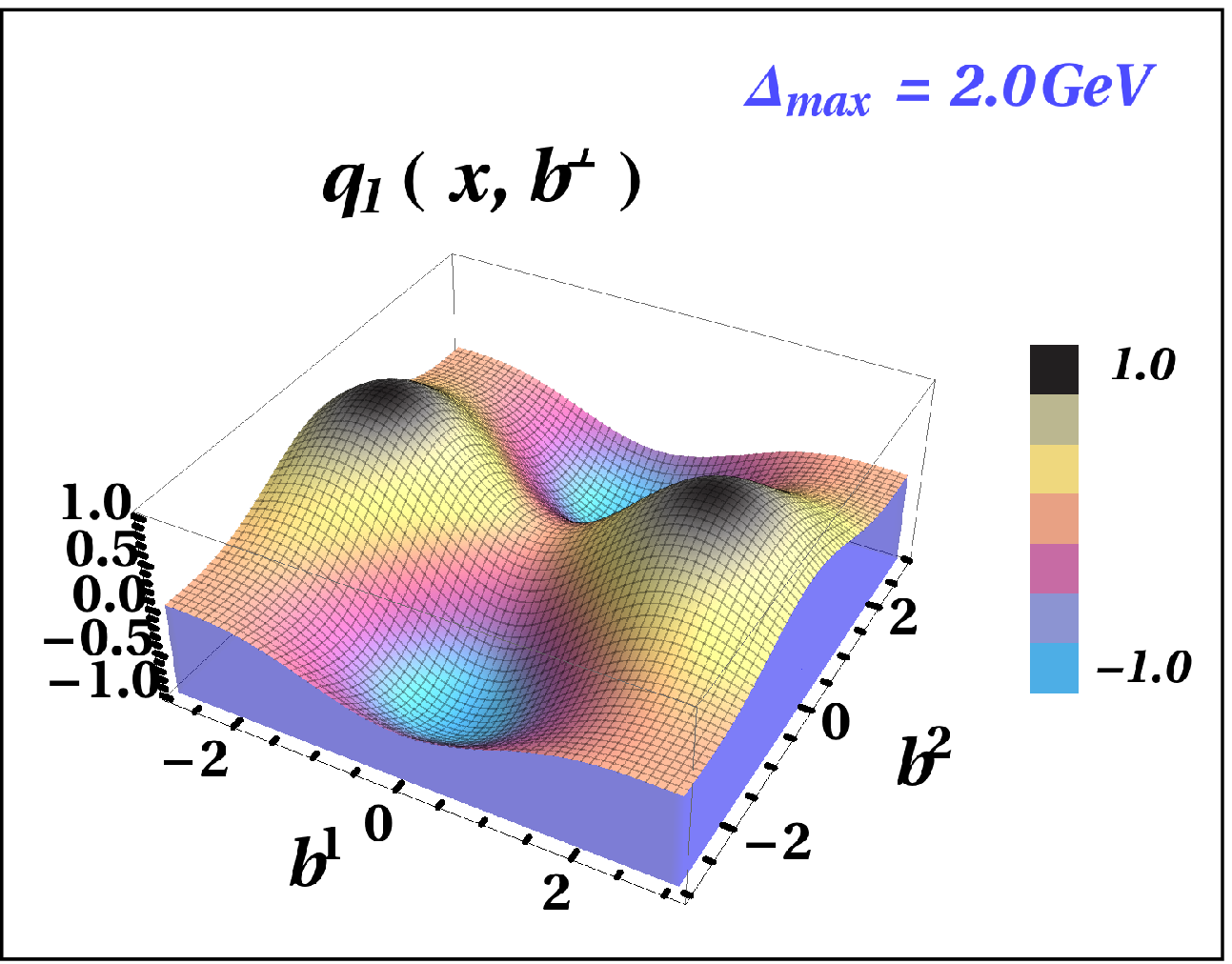}
\hspace{0.1cm}
\tiny{(b)}\includegraphics[width=7cm,height=6cm,clip]{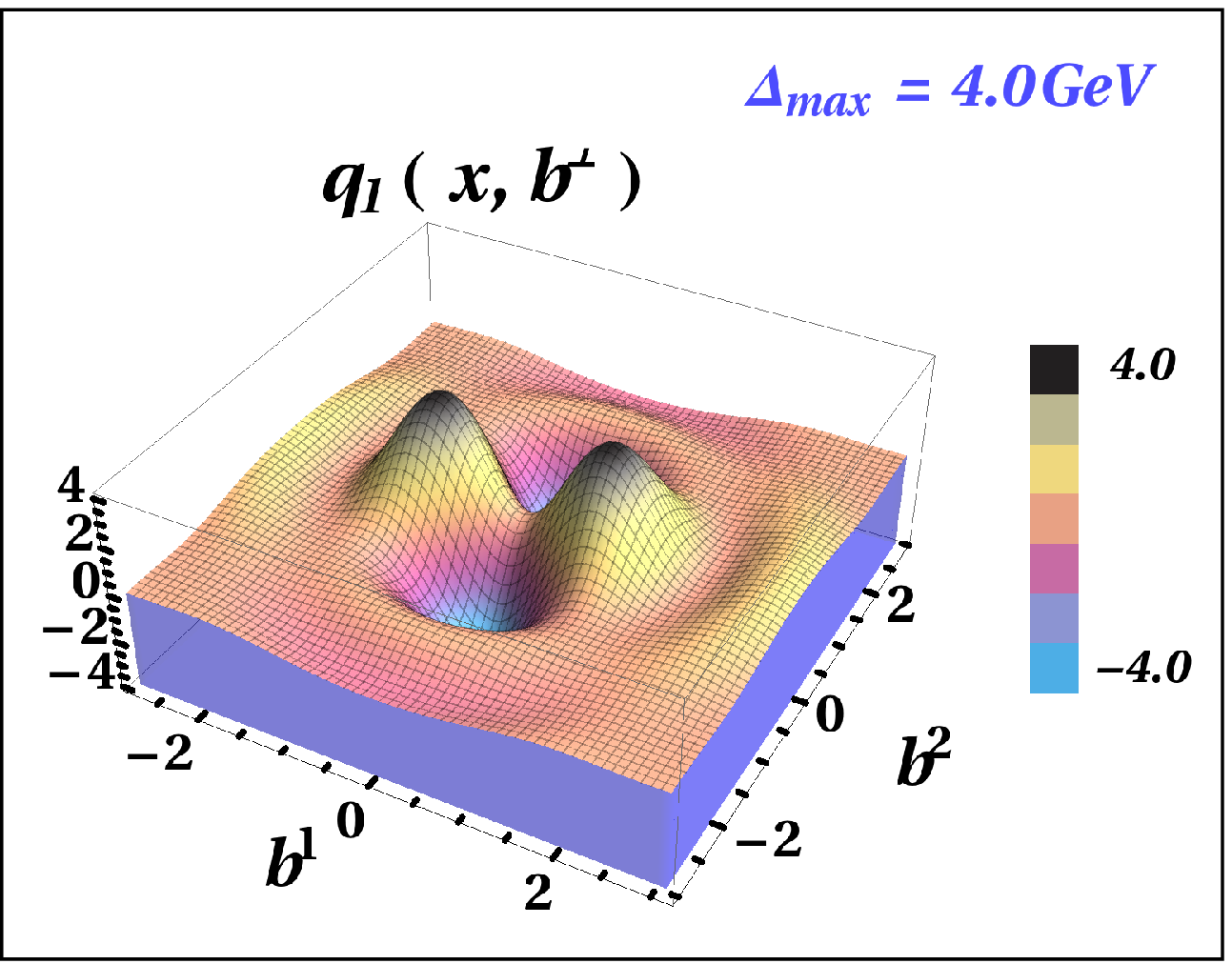}
\end{minipage}
\begin{minipage}[c]{0.99\textwidth}
\tiny{(c)}\includegraphics[width=7cm,height=6cm,clip]{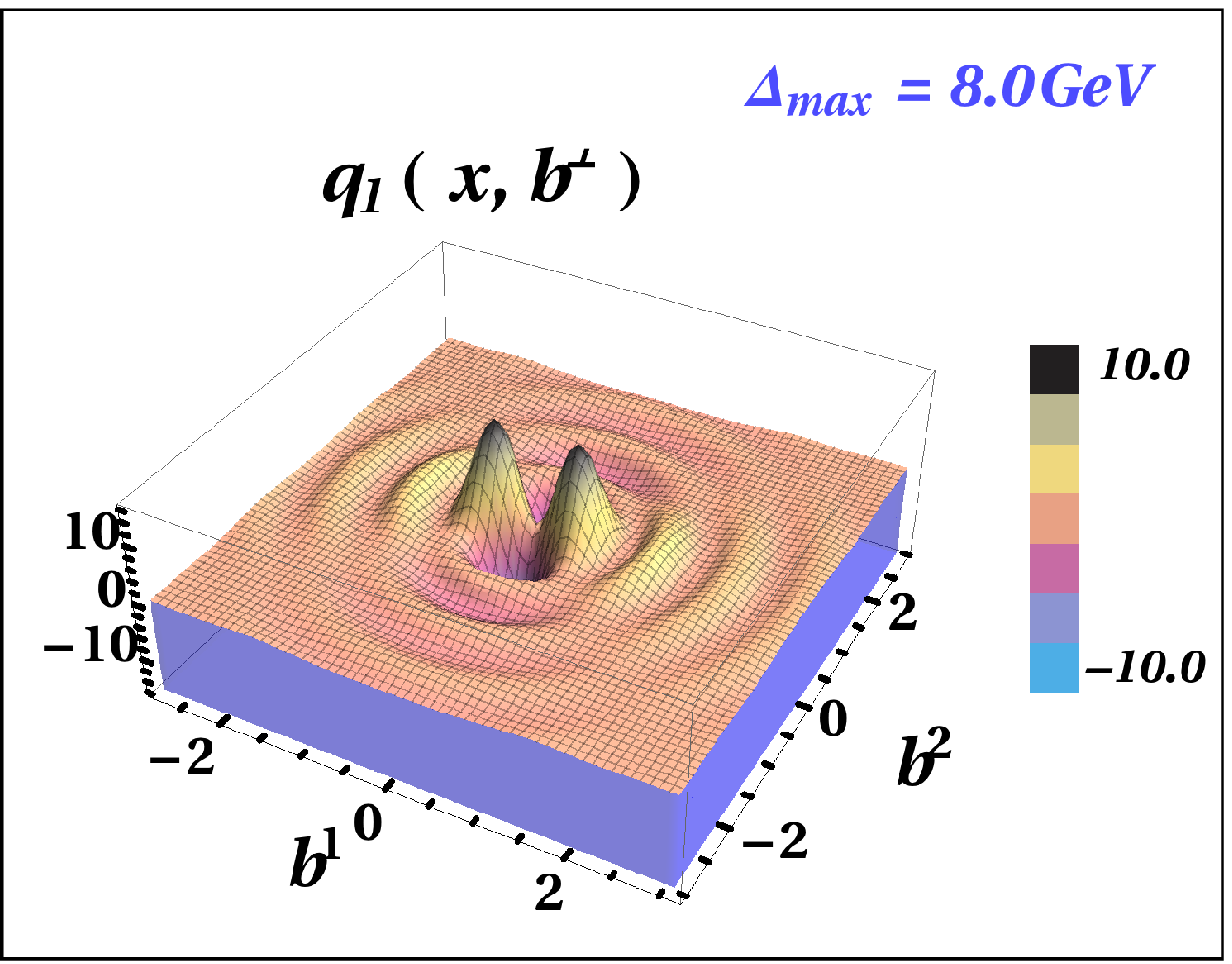}
\hspace{0.1cm}
\tiny{(d)}\includegraphics[width=7cm,height=6cm,clip]{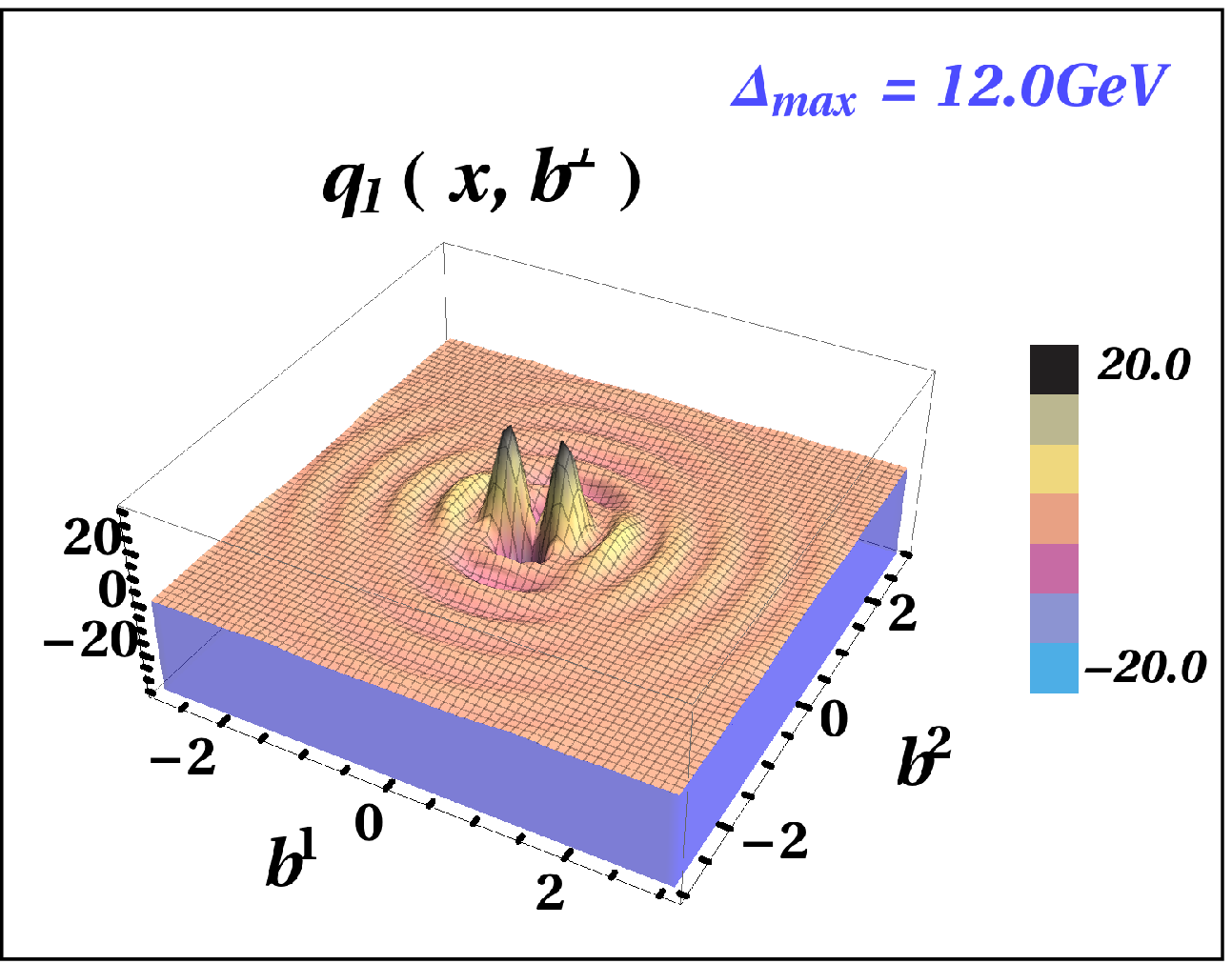}
\end{minipage}
\caption{\label{fig3}(Color online) Plots of  $q_1(x,b^\perp)$ vs. $b^1,b^2$ for
different values of $\Delta_{max}$. $b^1$ and $b^2$ are in
$\mathrm{GeV}^{-1}$ and  $\Delta_{max}$ is in GeV. $x=0.3$.} 
\end{figure}

\be
q_1(x,b^\perp)={1\over 4 \pi^2} \int d^2 \Delta^\perp e^{-i \Delta^\perp 
\cdot b^\perp} E_1 (x,\Delta^\perp);
\ee
where $t=-{(\Delta^\perp)}^2$ and $b^\perp$ is the transverse impact
parameter conjugate to $\Delta^\perp$. One then gets
\be
q_1(x,b^\perp)=\frac{1}{4\pi^2} \int d^2\Delta^\perp e^{-ib^\perp \cdot
\Delta^\perp} ({(\Delta^1)}^2-{(\Delta^2)}^2) f(x) Q(x,t),
\ee 
where 
\be
f(x)= \frac{\alpha e_{q}^2}{2 \pi}x (1-x)^{3}
\hspace{0.5cm}, ~~~~ \hspace{0.5cm} Q(x,t)=
\int_0^{1} \frac{dq}{B(q)}\hspace{0.2cm}
( (1-q)^2-(1-q));
\ee

This can be written as,

\be
q_1(x,b^\perp)&=&\frac{1}{4\pi^2} \Big(\frac{\partial^2}
{\partial {(b^2)}^2}-\frac{\partial^2}{\partial {(b^1)}^2}\Big) 
\int  d^2\Delta^\perp e^{-ib^\perp \cdot \Delta^\perp}  f(x) Q(x,t)
\nonumber\\&=&\frac{1}{2 \pi} 
 \Big(\frac{\partial^2}{\partial {(b^2)}^2}-\frac{\partial^2}
{\partial {(b^1)}^2}\Big)\Big[ \int_0^\infty  \Delta d\Delta
 \hspace{0.2cm}J_0(b\Delta)\hspace{0.2cm}
 f(x) Q(x,t) \Big].
\ee
Here $\Delta = \mid \Delta^\perp \mid$ and $b=\mid b \mid $. Using the
integral representation for the Bessel function $J_0(x)$, the above can be
written as, 

\be
q_1(x,b^\perp)=\frac{1}{2\pi} \Big(\frac{\partial^2}{\partial {(b^2)}^2}
-\frac{\partial^2}{\partial {(b^1)}^2}\Big) \Big[ \int_0^\infty  \Delta d\Delta 
\hspace{0.2cm}\frac{1}{\pi} \int_0^{\pi} \mathrm{cos}(b \Delta
- \mathrm{\mathrm{sin}}   
\theta) d\theta \hspace{0.2cm} f(x) Q(x,t) \Big] .
\label{q1}
\ee 

We then get
\be
q_1(x,b^\perp)=\frac{1}{2\pi} \int_0^\infty 
\Delta d\Delta \hspace{0.2cm}\frac{1}{\pi} \int_0^{\pi} (P_2(b,\Delta
,\theta)
-P_1(b,\Delta ,\theta) )d\theta  \hspace{0.2cm} f(x) Q(x,t);
 \ee 
where
\be
P_2(b,\Delta ,\theta)=
-\frac{1}{b^3}  \Delta \mathrm{sin} \theta \Big[
 {(b^2)}^2 b \Delta 
\mathrm{cos} \Big( b  
\Delta \mathrm{sin} \theta \Big) \mathrm{sin} \theta +\\ \nn {(b^1)}^2 
\mathrm{sin} \Big(b
\Delta \mathrm{sin} \theta \Big)\Big]\\ 
P_1(b,\Delta,\theta) = -\frac{1}{b^3}  \Delta
\mathrm{sin}
 \theta \Big[ {(b^1)}^2 b 
\Delta  \mathrm{cos} \Big( b \Delta \mathrm{sin}
\theta
 \Big) \mathrm{sin} \theta +\\ \nn  {(b^2)}^2 
\mathrm{sin} \Big( b \Delta \mathrm{sin} \theta
\Big)\Big].
\label{qq}
\ee

\begin{figure}
\begin{minipage}[c]{0.99\textwidth}
\tiny{(a)}\includegraphics[width=7cm,height=6cm,clip]{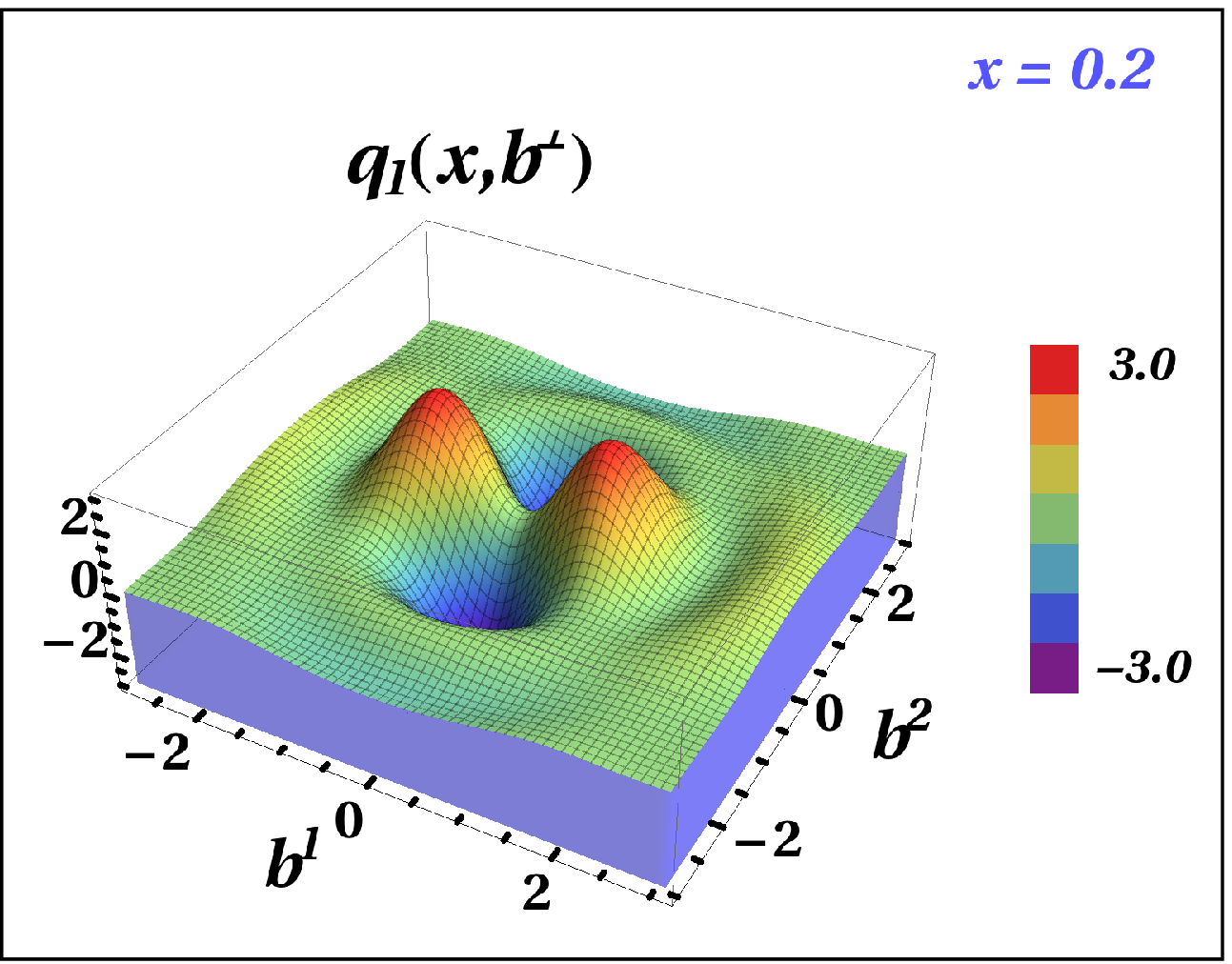}
\hspace{0.1cm}
\tiny{(b)}\includegraphics[width=7cm,height=6cm,clip]{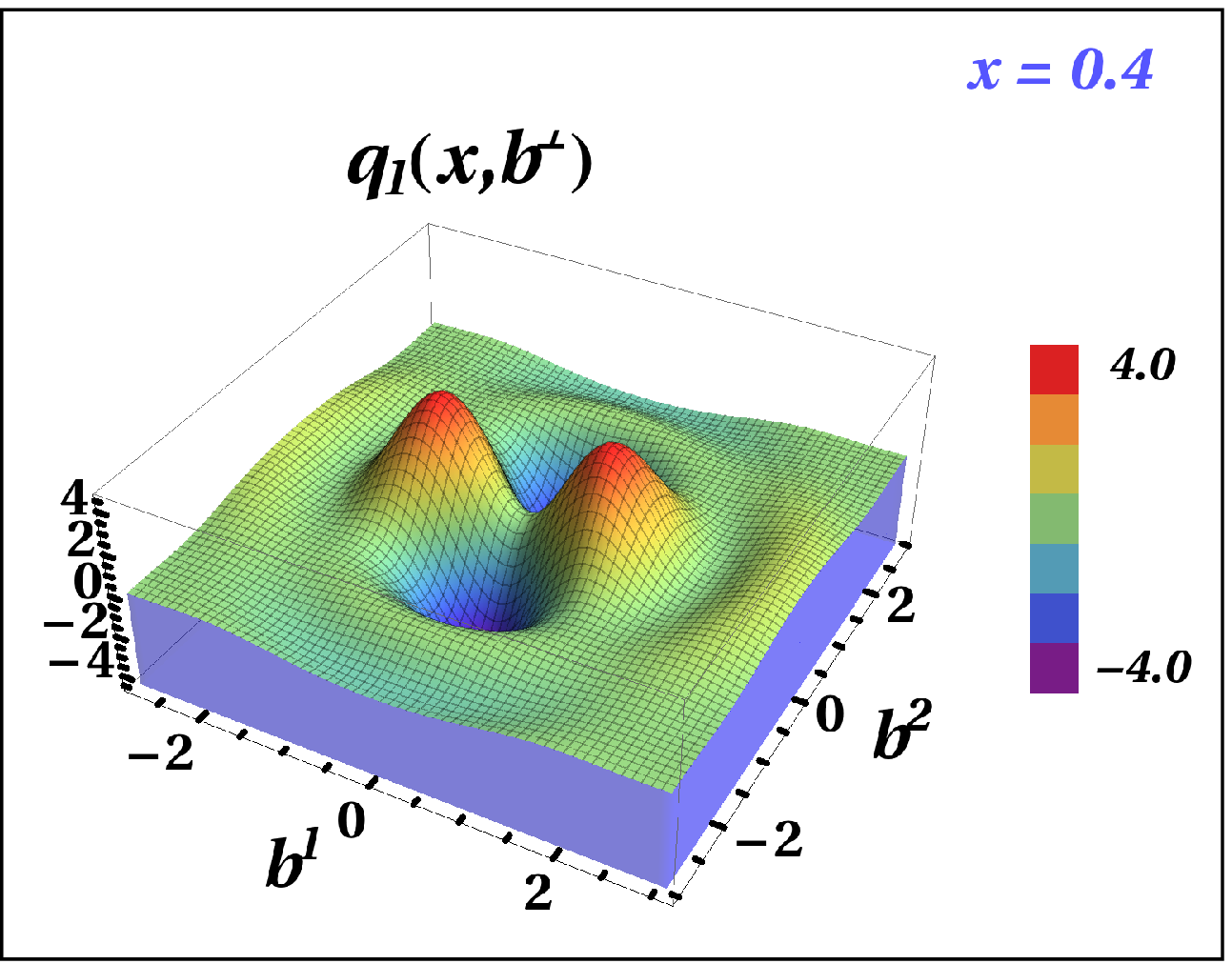}
\end{minipage}
\caption{\label{fig5}(Color online) Plots of $q_1 (x,b^\perp)$ vs $b^1,b^2$ for
different values of $x$ and at $\Delta_{max} = 4.0$ GeV.} 
\end{figure}

\vspace{0.2in}   

{\bf{Numerical Results}}
\vspace{0.2in}   

We next discuss the numerical results. In Fig. 1, we have shown the helicity
flip GPD of the photon as functions of $\Delta^\perp$. As we mentioned
before, we took $\zeta=0$. In this kinematical limit, the GPDs are
represented by overlaps of two-particle photon LFWFs. When $\zeta$ is
non-zero or the momentum transfer between the initial to final photon has a
longitudinal component, off-diagonal particle number changing overlaps of
LFWFs has to be considered as well \cite{overlap}. 
We take the mass of the quark and the
antiquark in the photon to be the same and equal to 3.3 MeV. As seen in Eq.
(\ref{lfwf}),
the GPDs have two contributions. When $x$ is positive, the contribution comes
from the active quark in the photon; and when $x$ is negative, the active
antiquark
contributes. In the numerical plots, we take $0<x<1$ and show the quark
contribution.  Fig. 1 shows the helicity flip GPD $E_1(x,\Delta^\perp)$ as 
functions of $\Delta^1$ and $\Delta^2$ and for different values of $x$ for fixed
values of $t$. $E_1(x,\Delta^\perp)$ is zero when $\Delta^1=\Delta^2$. The
curvature is sharper as $\mid t \mid$ decreases. As we already saw,
$E_1(x,\Delta^\perp)$ has a quadrupole structure in $\Delta^\perp$ plane coming from the
${(\Delta^1)}^2-{(\Delta^2)}^2$. Such quadrupole structure is due to the
spin flip of  a spin one particle and corresponds to an overlapping LFWF
with two units of orbital angular momentum. The structure can be contrasted
with the GPD $E(x,\Delta^\perp)$ of a spin $1/2$ composite particle 
like a dressed
electron or a proton \cite{quark,model,manohar2}. It is to be noted that the off-forward 
matrix element similar to Eq. (\ref{def}) for a proton target is parametrized in terms of the    
GPDs $H$ and $E$. When the final proton has the same helicity as the initial
proton, and $\zeta$ is non-zero, both the GPDs contribute. However when the    
helicity of the proton is flipped then only the GPD $E$ contributes. A
parametrization of the off-forward matrix element for a spin one massive
target like deuteron was given in \cite{one}. So far no such parametrization is
available for the photon GPDs and in this work as well as in two previous
publications \cite{ours1,ours2} we calculate the full off-forward matrix element 
using overlaps of photon LFWFs. In Fig. 2(a) we have plotted the helicity 
flip photon GPD $E_1(x,\Delta^\perp)$ vs. $x$ for different values of $t$
and a fixed value of $\phi=tan^{-1}{\Delta^2\over \Delta^1}$. As before, 
we have plotted for the region $0<x<1$, where the contribution to the 
GPD comes from the active quark. The peak of 
$E_1(x,\Delta^\perp)$ increases as $-t$ increases and also shifts towards
larger value of $x$. The GPD is zero both at $x=0$ and $x=1$. In fact, the
GPD is zero when $\Delta^\perp=0$. This is because in order to flip the
helicity one needs non-zero OAM in the two-particle LFWFs and the OAM is 
zero when there is no momentum transfer in the transverse direction. 
At $x=0$ and $x=1$ all momenta are carried by either the quark or the
antiquark in the photon. Then there is no relative motion and no OAM
contribution. In Fig.
2 (b) we have plotted the Fourier transform (FT) of the helicity-flip photon GPD,
$q_1(x,b^\perp)$ vs. $x$ for different $b=\mid b^\perp \mid$ at a fixed
value of $\beta=tan^{-1}{b^2\over b^1}$. $q_1(x,b^\perp)$       
is symmetric with respect to $x=0$ and $x=1$ and maximum when $x=0.5$, that
is when the quark and the antiquark carry equal momenta. As seen in Eq.
(\ref{q1}), 
$q_1(x,b^\perp)$ has a quadrupole structure, that comes because it involves
a helicity flip of a spin one object (photon). This quadrupole
structure is visible in the 3D plots of Figs 3 and 4. 
In the ideal definition of the Fourier transform,  the limits of the 
$\Delta^\perp$ integration should be from $0$ to $\infty$. As we saw for the 
photon GPDs that do not involve a helicity flip \cite{ours1,ours2}, 
the $\Delta^\perp$ independent terms then give a $\delta(b^\perp)$ in impact parameter space. For non-zero
$\Delta^\perp$, we get a smearing in $b^\perp$. For the GPD with helicity flip,
from Eq. (\ref{q1}), we see that it involves a distortion in $b^\perp$ space. The GPD as well 
as its FT  is zero when $\Delta^\perp=0$, which means that it is purely an effect 
of the orbital angular momentum of the LFWF. In the actual numerical calculation
we have imposed an upper limit on the $\Delta^\perp$ integration, denoted by $\Delta_{max}$.
Fig. 3 shows a plot of $q_1(x,b^\perp)$ vs. $b^1$ and $b^2$ for a fixed value of $x=0.3$ and different
 values of $\Delta_{max}$.
It is seen that as $\Delta_{max}$ increases the peaks become sharper, which means that
the distortion in $b^\perp$ space moves closer to the origin. Fig. 4 shows plots of
$q_1(x,b^\perp)$ vs. $b^1$ and $b^2$ for a  fixed value of $\Delta_{max}$ and two different values of $x$. 
The magnitude of the peaks depend on $x$.    

\vspace{0.2in}          
{\bf{Conclusion}}
\vspace{0.2in}          

In this work we have calculated the GPDs of the photon when the helicity of
the target photon is flipped. We expressed the GPDs in terms of overlaps of
photon LFWFs. In the kinematics when the momentum transfer between the
initial and the final photon is purely in the transverse direction, the GPDs
involve diagonal overlaps of two-particle LFWFs at leading order in the
electromagnetic coupling $\alpha$ and zeroth order in the strong coupling
$\alpha_s$.  Such two particle LFWFs of the photon can be calculated in
light-front Hamiltonian perturbation theory. Taking a Fourier transform of
the GPDs with respect to $\Delta^\perp$ we obtained impact parameter
dependent parton distributions. Like the proton GPD $E$ the helicity
flip GPD of the photon represents a distortion of the parton distribution in
the impact parameter space. This is due to the orbital angular momentum
contribution coming from the LFWFs. As photon is a spin one object, one
needs OAM of two units in the overlapping LFWF to flip the helicity. The
expected quadrupole structure is visible in impact parameter space. For the
proton, such distortion in $b^\perp$ space has been found to be related to
the Sivers function in some models. It will be interesting to check if such 
relations exist also for the photon. For this it is necessary to have a
parametrization of the off-forward as well as
the transverse momentum dependent  matrix elements for the photon.   

\vspace{0.2in}          

{\bf{Acknowledgments}}

\vspace{0.2in}          

This work is supported by the DST project SR/S2/HEP-029/2010, Govt. of
India. We thank B. Pire for suggesting this topic
and for helpful discussions.     


\end{document}